\documentclass[12pt]{article}
\usepackage{amsmath}
\usepackage{graphicx}
\usepackage{subfigure}
\usepackage{hyperref}

\begin{document}

\author{Ernst Trojan \and \textit{Moscow Institute of Physics and Technology} \and 
\textit{PO Box 3, Moscow, 125080, Russia}}
\title{Tachyonic Dirac sea}
\maketitle

\begin{abstract}
We consider a system of many fermions with tachyonic energy spectrum $%
\varepsilon _k=\sqrt{k^2-m^2}$ and clarify that tachyons with imaginary
energy and low momentum ($k<m$) play the role of Dirac sea in a many-tachyon
Fermi system and make contribution to the thermodynamical functions. The
energy and pressure acquire additional constant terms that, however, is not
reflected in the sound speed. Replacement $m\mapsto im$ results in the
thermodynamical functions and the sound speed of an ordinary Fermi gas. When
the Fermi momentum approaches the Dirac sea level $k_F\rightarrow m$, the
group velocity of most tachyons above the sea is unbound, while the sound
speed tends to infinity. This scenario is not encountered in practice
because the cold tachyon Fermi gas becomes unstable with respect to
hydrodynamical perturbations as soon as $k_F<\sqrt{3/2}m$. The particle
number density of a stable many-tachyon system is always finite and exceeds
the critical value depending on the tachyon mass $m$.
\end{abstract}

\section{Introduction}

Tachyons are instabilities of the field theory often considered in
cosmological models. The tachyonic Dirac equation \cite{D1,D2,D3,D4,D5} 
\begin{equation}
\left( i\gamma ^\mu \partial _\mu -\gamma _5m\right) \psi =0  \label{dir}
\end{equation}
has plane-wave solution 
\begin{equation}
\psi \sim \exp \left( i\vec k\cdot \vec r-i\varepsilon _kt\right)
\label{wav}
\end{equation}
with the energy spectrum 
\begin{equation}
\varepsilon _k=\sqrt{k^2-m^2}  \label{tah}
\end{equation}
that can be presented in the form 
\begin{equation}
\varepsilon _k=\mathrm{Re}\varepsilon _k+i\mathrm{Im}\varepsilon _k=\sqrt{%
k^2-m^2}\Theta \left( k-m\right) +i\sqrt{m^2-k^2}\Theta \left( m-k\right)
\label{com}
\end{equation}
where 
\begin{equation}
\Theta \left( x\right) =\left\{ 
\begin{array}{c}
1\qquad x\geq 0 \\ 
0\qquad x<0
\end{array}
\right.  \label{tet}
\end{equation}
is the Heaviside step function. It is clear that solution (\ref{wav}) at $%
\mathrm{Im}\varepsilon _k=0$ describes a stationary plane wave $\left\| \psi
\right\| =\mathrm{const}$, while \ 
\begin{equation}
\left\| \psi \right\| =\sqrt{\psi ^{\dagger }\psi }\sim \exp \left( -\mathrm{%
Im}\varepsilon _kt\right)  \label{mag}
\end{equation}
corresponds to an unstable particle at $\mathrm{Im}\varepsilon _k\neq 0$ ($%
k<m$).

A system of many tachyons can be studied in the frames of statistical
mechanics \cite{M84,DHR89,KRS07,TV2011c}. A single tachyon with imaginary
energy (\ref{com}) and small momentum ($k<m$) cannot be presented by a
stable plane wave (\ref{wav}) because the amplitude of its wave function is
subject to decay (\ref{mag}). How to operate with unstable sector $k<m$ when
a many-particle system is considered? Indeed, we can exclude unstable
particles from consideration at all and operate with the only stable
particles whose momentum $k\geq m$. However, a Fermi gas of many tachyons
may contain occupied states at low momentum $k<m$ if they satisfy the Pauli
exclusion principle. When we estimate the thermodynamical functions of a
cold tachyon Fermi gas, should we take into account or disqualify the
''unphysical'' states with imaginary energy ($k<m$)? This problem is
clarified in the present paper.

\section{Thermodynamical functions}

Starting with the thermodynamical potential \cite{K417}

\begin{equation}
\Omega =-\frac{\gamma T}{2\pi ^2}V\int\limits_0^\infty \ln \left( 1+\exp 
\frac{\mu -\varepsilon _k}T\right) k^2dk  \label{z}
\end{equation}
for a system of constant number of $N$ fermions we determine the Helmholtz
free energy $F=\Omega +\mu N$, the pressure 
\begin{eqnarray}
P=-\frac \Omega V=\frac \gamma {2\pi ^2}T\int\limits_0^\infty \ln \left(
1+\exp \frac{\mu -\varepsilon _k}T\right) k^2dk=\qquad \qquad \qquad  && 
\nonumber \\
=\frac{\gamma T}{6\pi ^2}\left. k^3\ln \left( 1+\exp \frac{\mu -\varepsilon
_k}T\right) \right| _0^\infty +\frac \gamma {6\pi ^2}\int\limits_0^\infty
f_{\varepsilon \,}\frac{d\varepsilon _k}{dk}\,k^3dk &&  \label{px}
\end{eqnarray}
the energy density 
\begin{equation}
E=-\frac{T^2}V\frac{\partial \left( F/T\right) _{V,\mu }}{\partial T}=\frac
\gamma {2\pi ^2}\int\limits_0^\infty \,\!f_\varepsilon \,\varepsilon
_{k\,}k^2dk  \label{e}
\end{equation}
and the particle number density 
\begin{equation}
n=\frac NV=-\frac 1V\left( \frac{\partial \Omega }{\partial \mu }\right)
_{V,T}=\frac 1V\frac{\partial \left( T\ln Z\right) _{V,T}}{\partial \mu }%
=\frac \gamma {2\pi ^2}\int\limits_0^\infty f_{\varepsilon \,}\,k^2dk
\label{n}
\end{equation}
where the Fermi-Dirac distribution function is 
\begin{equation}
f_\varepsilon =\frac 1{\exp \left[ \left( \varepsilon _k-\mu \right)
/T\right] +1}  \label{f}
\end{equation}
and the chemical potential $\mu $ satisfies relation 
\begin{equation}
\mu =\left( \frac{\partial F}{\partial N}\right) _{T,V}  \label{mu0}
\end{equation}
Formulas (\ref{px})-(\ref{n}) at zero temperature yield the third law of
thermodynamics (Nernst heat theorem)

\begin{equation}
E+P=\mu n  \label{ner}
\end{equation}
while formula (\ref{mu0}) is reduced to $\mu =dE/dn$, and the sound speed at
zero temperature is determined by formula 
\begin{equation}
c_s^2=\frac{dP}{dE}=\frac n\mu \frac{d\mu }{dn}  \label{c0}
\end{equation}

Formula (\ref{px}) is reduced to 
\begin{equation}
P=\frac \gamma {6\pi ^2}\int\limits_0^\infty f_{\varepsilon \,}\frac{%
d\varepsilon _k}{dk}\,k^3dk  \label{p}
\end{equation}
under assumption 
\begin{equation}
\lim_{k\rightarrow \infty }\varepsilon _k=+\infty \qquad \lim_{k\rightarrow
0}k^3\varepsilon _k=0  \label{ass}
\end{equation}
\textrm{\ }Most particles and quasi-particles satisfy condition (\ref{ass}),
particularly, it is so for the tachyonic energy spectrum (\ref{tah}).

\section{Cold tachyon Fermi gas}

We may expect that the chemical potential is a complex quantity $\mu =%
\mathrm{Re}\mu +i\mathrm{Im}\mu $. If we expect that $\mathrm{Re}\mu >0$ and 
$\mathrm{Im}\mu =0$, then, the distribution function (\ref{f}) at zero
temperature is reduced to 
\begin{equation}
f_\varepsilon =\Theta \left( \varepsilon _F-\varepsilon _k\right)  \label{f3}
\end{equation}
that is equivalent to 
\begin{equation}
f_\varepsilon =\Theta \left( k_F-k\right)  \label{f33}
\end{equation}
where 
\begin{equation}
\varepsilon _F=\mathrm{Re}\mu |_{T=0}  \label{ef}
\end{equation}
is the Fermi energy corresponding to the Fermi momentum $k_F$ according to
relation 
\begin{equation}
\varepsilon _F=\sqrt{k_F^2-m^2}  \label{ef1}
\end{equation}

Substituting the distribution function at zero temperature (\ref{f33}) in (%
\ref{n}), we have

\begin{equation}
n=\frac \gamma {2\pi ^2}\int\limits_0^{k_F}k^2dk=\frac{\gamma k_F^3}{6\pi ^2}%
=\frac{\gamma \sqrt{\left( \varepsilon _F^2+m^2\right) ^3}}{6\pi ^2}
\label{n3}
\end{equation}
that at $\varepsilon _F\rightarrow 0$ tends to 
\begin{equation}
n\rightarrow n_{\star }=\frac{\gamma m^3}{6\pi ^2}  \label{nsr}
\end{equation}

According to (\ref{c0}) and (\ref{n3}) we find the sound speed$\allowbreak $%
\begin{equation}
c_s^2=\frac 13\frac{k_F^2}{k_F^2-m^2}  \label{c}
\end{equation}
that is superluminal when $k_F<\sqrt{3/2}m$ corresponds to 
\begin{equation}
n<n_T=\sqrt{\frac{27}8}n_{\star }\simeq 1.84n_{\star }  \label{ca1}
\end{equation}

Substituting the tachyonic energy spectrum (\ref{com}) and the distribution
function (\ref{f33}) in (\ref{e}) and in (\ref{p}), we find the energy
density and pressure at zero temperature 
\begin{equation}
E=\frac \gamma {2\pi ^2}\int\limits_m^{k_F}\,k^{2\,}\mathrm{Re}\varepsilon
_k\,dk+\frac{i\gamma }{2\pi ^2}\int\limits_0^m\,k^{2\,}\mathrm{Im}%
\varepsilon _k\,dk  \label{e1}
\end{equation}

\begin{equation}
P=\frac \gamma {6\pi ^2}\int\limits_m^{k_F}k^{3\,}\frac{d\mathrm{Re}%
\varepsilon _k}{dk}\,dk+\frac{i\gamma }{6\pi ^2}\int\limits_0^mk^{3\,}\frac{d%
\mathrm{Im}\varepsilon _k}{dk}\,dk  \label{p1}
\end{equation}
Imaginary constants 
\begin{equation}
E_0=\frac{i\gamma }{2\pi ^2}\int\limits_0^mk^{2\,}\mathrm{Im}\varepsilon
_k\,dk=\frac{i\gamma }{2\pi ^2}\int\limits_0^m\,k^2\sqrt{m^2-k^2}dk=\frac{%
i\gamma m^4}{32\pi }\qquad  \label{e0}
\end{equation}
\begin{equation}
P_0=\frac{i\gamma }{6\pi ^2}\int\limits_0^mk^{3\,}\frac{d\mathrm{Im}%
\varepsilon _k}{dk}\,dk=-\frac{i\gamma }{2\pi ^2}\int\limits_0^m\,\frac{k^4}{%
\sqrt{m^2-k^2}}\,dk=-\frac{i\gamma m^4}{32\pi }  \label{p0}
\end{equation}
are included in the energy density (\ref{e1}) and pressure (\ref{p1}),
playing the role of zero-point levels. Notable that the multiplier $5/(32\pi
)$ appears in the imaginary part of the vacuum energy of tachyonic modes in
the Chern-Simons electrodynamics \cite{A98}

We can exclude the 'vacuum' imaginary terms, making renormalization 
\begin{equation}
\bar E\mapsto E-E_0  \label{e00}
\end{equation}
\begin{equation}
\bar P\mapsto P-P_0  \label{p00}
\end{equation}
The real parts of energy density (\ref{e1}) and pressure (\ref{p1}) were
determined in \cite{TV2011c}: 
\begin{eqnarray}
\bar E=\frac \gamma {2\pi ^2}\int\limits_m^{k_F}k^{2\,}\mathrm{Re}%
\varepsilon _k\,dk=\frac \gamma {2\pi ^2}\int\limits_m^k\,k^{2\,}\sqrt{%
k^2-m^2}dk=\qquad \qquad \qquad \qquad &&  \nonumber \\
=\frac \gamma {8\pi ^2}k_{F\,}^3\varepsilon _F-\frac \gamma {16\pi
^2}m^2\left( k_{F\,}\varepsilon _F+m^2\ln \frac{k_F+\varepsilon _F}m\right)
\qquad &&  \label{e1r}
\end{eqnarray}
\begin{eqnarray}
\bar P=\frac \gamma {6\pi ^2}\int\limits_m^{k_F}k^{3\,}\frac{d\mathrm{Re}%
\varepsilon _k}{dk}dk=\frac \gamma {6\pi ^2}\int\limits_m^{k_F}\,\frac{k^4}{%
\sqrt{k^2-m^2}}dk=\qquad \qquad \qquad \qquad &&  \nonumber \\
=\frac \gamma {24\pi ^2}k_{F\,}^3\varepsilon _F+\frac \gamma {16\pi
^2}m^2\left( k_{F\,}\varepsilon _F+m^2\ln \frac{k_F+\varepsilon _F}m\right)
\qquad &&  \label{p1r}
\end{eqnarray}
and it is also enough for calculation of the sound speed $c_s^2=dP/dE$ that
coincides with (\ref{c}) because constants $E_0$ and $P_0$ in (\ref{e00})- (%
\ref{p00}) do not influence the result of differentiation.

\section{Alternative view}

One should not hesitate to consider the tachyonic energy spectrum (\ref{tah}%
) in the whole range of momentum $k$ where the energy includes imaginary
part (\ref{com}), in fact, tachyonic states with complex energy are known in
the solid state physics \cite{G2009}. The fact that we determine the right
energy density (\ref{e1}) and pressure (\ref{p1}) is immediately checked by
formal replacement $m\mapsto im$\ that yields the thermodynamical functions
of an ordinary relativistic Fermi gas 
\begin{equation}
E=\frac \gamma {2\pi ^2}\int\limits_0^{k_F}\,k^2\sqrt{k^2+m^2}dk=\frac
\gamma {8\pi ^2}k_{F\,}^3\varepsilon _F+\frac \gamma {16\pi ^2}m^2\left(
k_{F\,}\varepsilon _F-m^2\ln \frac{k_F+\varepsilon _F}m\right)  \label{e1o}
\end{equation}
\begin{equation}
P=\frac \gamma {2\pi ^2}\int\limits_0^{k_F}\,k^2\sqrt{k^2+m^2}dk=\frac
\gamma {24\pi ^2}k_{F\,}^3\varepsilon _F-\frac \gamma {16\pi ^2}m^2\left(
k_{F\,}\varepsilon _F-m^2\ln \frac{k_F+\varepsilon _F}m\right)  \label{p1o}
\end{equation}
constituted of subluminal particles with the energy spectrum $\varepsilon _k=%
\sqrt{k^2+m^2}$. The same conversion is applied to the sound speed (\ref{c}).

It should be also noted that the tachyonic sound speed (\ref{c}) becomes
infinite when $k_F\rightarrow m$ that corresponds to infinite group velocity 
\begin{equation}
v=\frac{d\varepsilon _k}{dk}=\frac k{\sqrt{k^2-m^2}}\rightarrow \infty
\label{inf}
\end{equation}
for all stable tachyons with real energy ($k>m$).

If the integration is truncated in the very beginning within the range $k\in
\left( m,\infty \right) $, then, the energy density (\ref{e}) remains the
same (\ref{e1r}), while the pressure (\ref{p}) will be 
\begin{eqnarray}
P=\frac \gamma {2\pi ^2}T\int\limits_m^{k_F}k^2\ln \left( 1+\exp \frac{%
\varepsilon _F-\varepsilon _k}T\right) dk=\qquad \qquad \qquad \qquad && 
\nonumber \\
=\frac \gamma {24\pi ^2}k_{F\,}^3\varepsilon _F+\frac \gamma {16\pi
^2}m^2\left( k_F\varepsilon _F+m^2\ln \frac{k_F+\varepsilon _F}m\right) -%
\frac{\gamma m^3}{6\pi ^2}\varepsilon _F &&  \label{p1m}
\end{eqnarray}
Now conversion to the ordinary Fermi gas (\ref{e1o})-(\ref{p1o}) by
replacement $m\mapsto im$ is not working. The relevant sound speed \cite
{T2011j,B2011} 
\begin{equation}
c_s^2=\frac{dP}{dE}=\frac 13\frac{k_F^2+mk_F+m^2}{k_F^2+mk_F}  \label{cm}
\end{equation}
is finite at any $k_F$ that seems unrealistic because all tachyons of the
thermodynamical ensemble are traveling at infinite high velocity (\ref{inf})
when $k_F\rightarrow m$. All this implies that unstable sector $k<m$ should
not be omitted when we analyze the properties of a many-tachyon Fermi system.

\section{Conclusion}

Although a single tachyon has real energy (\ref{tah}) at large momentum $k>m$%
, a system of tachyons in a thermodynamical ensemble obeys the Fermi-Dirac
statistics and unstable states with imaginary energy ($k<m$) should be
included in analysis.

The properties of a cold tachyon Fermi gas depend on its density. The medium
is stable with respect to hydrodynamical perturbations only at high density $%
n\geq n_T$ (\ref{e1r}). When we consider a stable system of many fermions
with tachyonic energy spectrum (\ref{tah}), the quantum states $k<m$\ are
responsible for imaginary constants (\ref{e0}) and (\ref{p0}) added to the
thermodynamical functions (\ref{e1}) and (\ref{p1}). It implies reset of the
zero-point levels of the energy density (\ref{e00}) and pressure (\ref{p00}%
), however, the sound speed (\ref{c}) is independent on them.

\begin{figure}[tbp]
\caption{Graphical illustration of the Dirac sea of massive subluminal
particles (a), massless particles (b) and tachyons (c). }
\label{sea}{\includegraphics[scale=0.35]{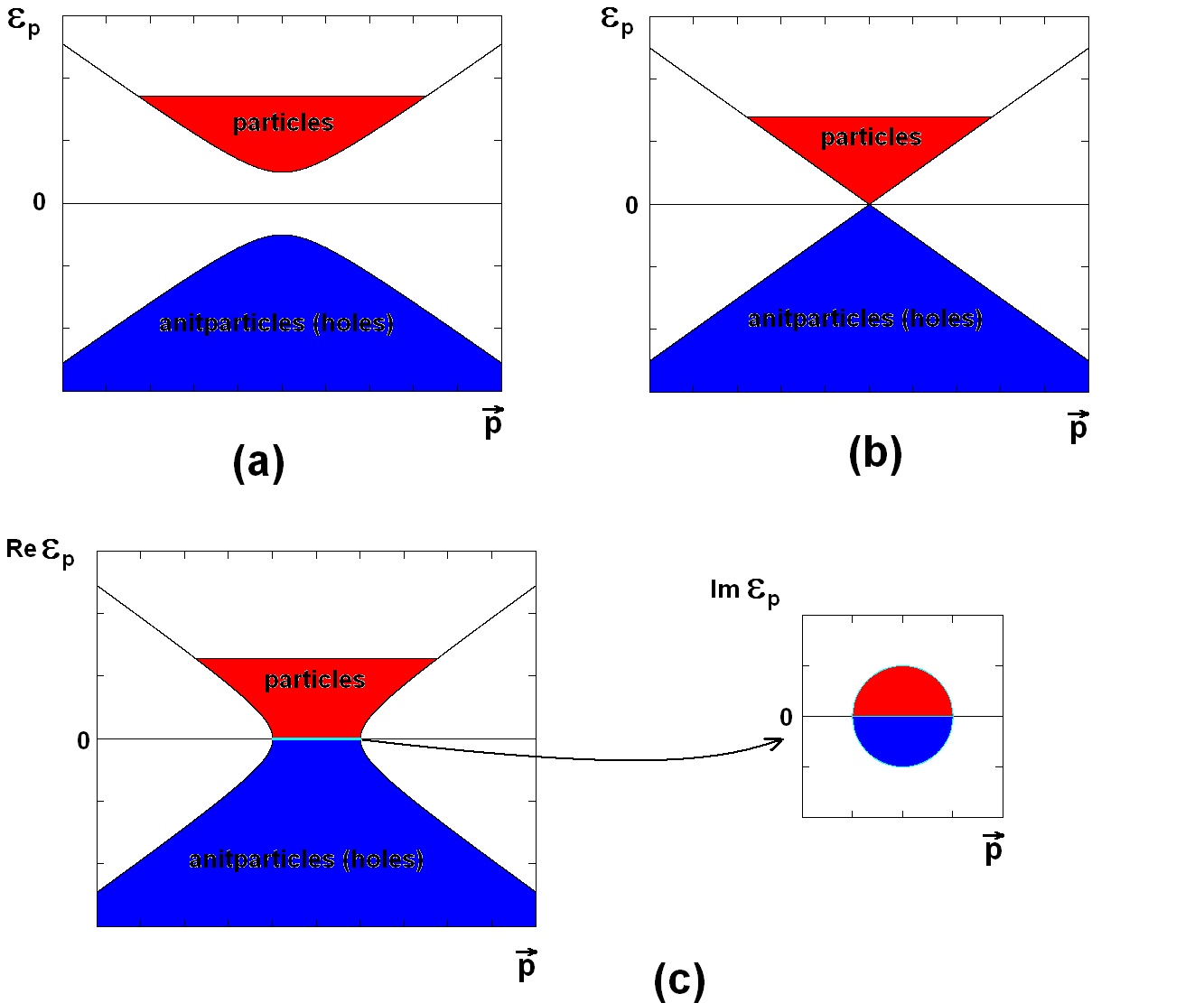}}
\end{figure}

The tachyons with imaginary energy and small momentum $k<m$ are inherent in
any many-tachyon Fermi system and we cam imagine them in the tachyonic Dirac
sea (see Fig.~\ref{sea}) because all tachyons from the whole sea ($k<m$) are
making contribution to the thermodynamical functions even when the Fermi
momentum itself is above the sea level $k_F>m$. Operating with the only real
part of tachyonic energy spectrum (\ref{com}), we should be careful to
complete this mathematical trick for determining the real part of the energy
density (\ref{e1r}) and pressure (\ref{p1r}), and, hence, again coming to
the sound speed (\ref{c}) \cite{TV2011c}. However, if we exclude the
''unphysical'' particles from the sea ($k<m$), we come to the tachyonic
thermodynamical functions (\ref{e1r}) and (\ref{p1m}) that are not
convertible to the thermodynamical functions of an ordinary Fermi gas (\ref
{e1o})-(\ref{p1o}) after replacement $m\mapsto im$, while the relevant sound
speed (\ref{cm}) remains finite even at $k_F\rightarrow m$ when most
''physical'' tachyons ($k>m$) travel at infinite high velocity (\ref{cm}).

Now we have seen the validity of formulas for the energy density (\ref{e1}),
pressure (\ref{p1}), sound speed (\ref{c}). It is important that the
causality condition $c_s^2\leq 1$ is satisfied at high density $n>n_T$ (\ref
{ca1}) which is definitely higher than $n_{\star }$ (\ref{nsr})
corresponding to $k_F=m$. The properties of a cold tachyon Fermi gas at
small density $n<n_{\star }$ may attract only pure theoretical interest
because the system becomes unstable as soon as its density decreases below $%
n=n_T\simeq 1.84n_{\star }$, and applied problems, like the tachyon stars 
\cite{T2011k}, require the only knowledge of stable tachyon Fermi gas.

The author is grateful to Erwin Schmidt for discussions.

\newpage 

\end{document}